\documentstyle[aps,prbbib,twocolumn,epsfig]{revtex}
\begin{document}
\draft

\twocolumn[\hsize\textwidth\columnwidth\hsize\csname@twocolumnfalse\endcsname

\title{Dephasing Effect in Photon-Assisted Resonant 
Tunneling through Quantum Dots}

\author{Junren Shi$^1$, Zhongshui Ma$^{1,2}$ and X.C. Xie$^{1}$}

\address{\protect\( ^{1}\protect \) Department of Physics, 
Oklahoma State University, Stillwater, OK74078\\
and\\
\protect\( ^{2}\protect \)Advanced Research Center, Zhongshan University,
Guangzhou, China}

\maketitle
\begin{abstract}
We analyze dephasing in single and double quantum dot systems. The decoherence
is introduced by the B\"{u}ttiker model with current conserving fictitious
voltage leads connected to the dots. By using the non-equilibrium Green 
function method, we investigate the dephasing effect on the tunneling
current. It is shown that a finite dephasing rate leads to observable
effects. The result can be used to measure
dephasing rates in quantum dots. 
\end{abstract}
\pacs{73.23.Ad, 73.40.Gk, 72.10.Bg} \medskip]

Recently, several
experimental\cite{oosterkamp,schedelbeck,tsui}
and theoretical\cite{stoof,wang} studies have been
devoted to
the analysis of effects of a time-dependent field on the resonance
tunneling through coupled double quantum dots.
Quantum dot systems are of great current interests because of their
fundamental physics as well as potential applications as possible
quantum-computing devices.
For the purpose of quantum computation, phase coherence plays an
important role. 
However, in the previous studies of
double-dot systems, the dephasing effect, caused by the
electron-electron or electron-phonon interaction, has been ignored.

In this paper we study the decoherence
effect on I-V characters 
in single and double dot systems. In a single-dot system,
we find that dephasing causes minor changes in the tunneling current.
On the other hand, in the pumping set-up of a double-dot system,
in which the chemical potentials on the left and right measurement
leads are equal, we find that
the photon-assisted resonant
tunneling current is sensitive to the dephasing rate. Thus, it provides
a possible way to measure the dephasing rate in double-dot systems.

To introduce the dephasing effect into the system, we use the B\"{u}ttiker
model\cite{buttiker1,buttiker2}.
In this approach, a system is connected
to virtual electron reservoirs
through fictitious voltage leads. With certain possibility, electrons
are scattered into these reservoirs, lose their phase memories,
then re-injected into the system. The chemical potential of the reservoir is
chosen such that there is no net current flow 
between the system and the reservoir in order to satisfy the
current conservation in the system.
In the
original B\"{u}ttiker model, only {\it dc} transport has been considered.
In our case, the gate voltage is modulated by injected microwave. Consequently,
we have to extend the constraint such that each Fourier component of
the current through a fictitious voltage lead vanishes.

\textbf{\underbar{Single Quantum Dot}}. 
Before discussing the more interesting case
of double-dot systems, we first consider the simpler system of a single
quantum dot. The dot is connected to two reservoirs
(\( l \) and \( r \)) and a fictitious 
voltage probe (\( \phi  \)). The potential
on the dot is controlled by a side-gate voltage. The Hamiltonian can be written
as 
\begin{eqnarray*}
H & = & H_{0}+H_{\phi },\\
\end{eqnarray*}
where
\begin{eqnarray*}
H_{0} & = & \sum _{\alpha k}\epsilon _{\alpha
k}(t)c^{+}_{\alpha k}
c_{\alpha k}+\epsilon (t)d^{+}d+\sum _{\alpha k}
[t_{\alpha k}c^{+}_{\alpha k}d+h.c.],\\
\end{eqnarray*}
and
\begin{eqnarray*} 
H_{\phi } & = & \sum _{k}\epsilon _{k}(t)c^{+}_{\phi k}c_{\phi k}
+\sum _{k}[t_{\phi k}c^{+}_{\phi k}d+h.c.].
\end{eqnarray*}
\( H_{0} \) is the Hamiltonian of the system without the fictitious probe and
\( \alpha =l,r \) are the indices for the left and right leads.
\( c^{+}_{\alpha k} \)(\( c_{\alpha k} \))
creates (annihilates) an electron in lead \( \alpha \)
while \( d^{+} \)(\( d \)) creates (annihilates) 
an electron in the dot. \( H_{\phi } \)
is the Hamiltonian of the fictitious probe, and 
\( c^{+}_{\phi k}(c_{\phi k}) \) is the electron
creation (annihilation) operator in the fictitious probe.
The microwave injection is modeled as {\it ac} 
side-bias which imposes a time-dependent
site energy of the quantum 
dot \( \epsilon (t)=\epsilon _{0}+ev_{ac}\cos \omega _{0}t \).
The chemical potential and the time-dependent voltage of the fictitious probe
are determined by the condition 
that the total current flowing through the probe
\( \phi  \) vanishes.

The time-dependent current from reservoir $\alpha$ to 
the dot can be expressed as\cite{wingreen}
\begin{eqnarray}
I_{\alpha }(t) & = & -\frac{2e}{\hbar }\int^{\infty }_{-\infty }
dt_{1}\int \frac{d\epsilon }{2\pi }\mathrm{Im}\left\{ e^{i\epsilon 
(t-t_{1})}\Gamma _{\alpha }(\epsilon )\right. \nonumber \\
 & \times  & \left. \exp \left[ i\int ^{t}_{t_{1}}d\tau 
\Delta_{\alpha }(\tau )\right] \left[ G^{<}(t,t_1 )+f_{\alpha }
(\epsilon )G^{r}(t,t_1 )\right] \right\} ,\label{Ialpha} 
\end{eqnarray}
where \( \Gamma _{\alpha }(\epsilon )=2\pi \rho _{\alpha }(\epsilon )
|t_{\alpha }|^{2} \) and \( \rho _{\alpha } \) is
the density of states of the reservoir \( \alpha  \).
$\alpha$ is the index for the left, right, or fictitious voltage lead.
For simplicity, we use the wide-band approximation, \textit{i.e.}, treating
the \( \Gamma _{\alpha }(\epsilon )\) as a constant,
\( \Gamma _{\alpha }(\epsilon )=\Gamma _{\alpha } \).
Within the wide-band approximation, the retarded Green function \( G^{r}
\) takes the form\cite{wingreen}
\begin{eqnarray*}
G^{r}(t,t') & = & -i\theta (t-t')\exp \left\{ -i\left[ \epsilon _{0}
-i\frac{\Gamma _{0}+\Gamma _{\phi }}{2}\right] (t-t')\right. \\
 &  & \left. -i\int ^{t}_{t'}d\tau \Delta _{ac}(\tau )\right\}, 
\end{eqnarray*}
where \( \Delta _{ac}(\tau )=ev_{ac}\cos \omega _{0}\tau  \), and 
\( \Gamma _{0}=\Gamma _{l}+\Gamma _{r} \)
is the energy broadening of the quantum dot due to the left and right leads,
and \( \Gamma _{\phi } \) is the broadening due to the fictitious voltage lead.
The spectral density, which relates to $G^r$ through
\begin{equation}
A_\alpha (\epsilon,
t)=\int^t_{-\infty}dt_1e^{i\epsilon(t-t_1)}e^{i\int^t_{t_1}d\tau
\bigtriangleup_{ac}(\tau)}G^r(t,t_1),
\end{equation}
is given by 
\[A_{\alpha }(\epsilon ,t)=
\sum ^{\infty }_{m=-\infty }\frac{J_{m}(\tilde{u}_{\alpha })}
{\tilde{\epsilon }^{(\phi )}_{m}(\epsilon )}e^{-i\tilde{u}_{\alpha }
\sin \omega _{0}t}e^{im\omega _{0}t},\]
where \( \tilde{\epsilon }^{(\phi )}_{m}(\epsilon )=\epsilon -
\epsilon _{0}-m\hbar \omega _{0}+i(\Gamma _{0}+\Gamma _{\phi })/2 \),
\( \tilde{u}_{0}=ev_{ac}/\omega _{0} \), 
and \( \tilde{u}_{\phi }=e(v_{\phi }-v_{ac})/\omega _{0} \). 

The Keldysh Green function is related 
to the retarded and advanced Green functions
through the Keldysh equation,
\[
G^{<}(t,t')=\int dt_{1}\int dt_{2}G^{r}(t,t_{1})\Sigma ^{<}(t_{1},t_{2})
G^{a}(t_{2},t').\]
Again, using the wide-band approximation, we obtain the Keldysh self-energy
\[
\Sigma ^{<}(t_{1},t_{2})=i[\sum _{\alpha }f_{\alpha }(\epsilon )
\Gamma _{\alpha }+f_{\phi }(\epsilon )\Gamma _{\phi }]\delta (t_{1}-t_{2}).\]
The Keldysh Green function is related to \( A_{\alpha }(\epsilon ,t) \) through
\begin{eqnarray*}
 &  & 2\int ^{t}_{-\infty }dt_{1}{\mathrm{Im}}\left
\{ e^{i\epsilon (t-t_{1})}\exp \left[ i\int ^{t}_{t_{1}}d\tau 
\Delta _{\alpha }(\tau )\right] G^{<}(t,t_{1})\right\} =\\
 &  & \sum _{\alpha }f_{\alpha }(\epsilon )
\Gamma _{\alpha }|A_{\alpha }(\epsilon ,t)|^{2}.
\end{eqnarray*}
After some algebra, the current flowing 
through the fictitious probe is found to be
\begin{equation}
\label{Iphi}
I_{\phi }(t)=-\frac{e}{\hbar }\Gamma _{\phi }\sum ^{\infty }_{m,n
=-\infty }\int \frac{d\epsilon }{2\pi }\frac{e^{i(m-n)\omega _{0}t}}
{\tilde{\epsilon }^{(\phi )}_{m}(\epsilon )[\tilde{\epsilon }^{(\phi )}_{n}
(\epsilon )]^{*}}{\mathcal{I}}_{mn}^{(\phi )}(\epsilon ),
\end{equation}
where
\begin{eqnarray*}
{\mathcal{I}}^{(\phi )}_{mn}(\epsilon ) & = & f_{\phi }(\epsilon )
J_{m}(\tilde{u}_{\phi })J_{n}(\tilde{u}_{\phi })[-i(m-n)\omega _{0})
-\Gamma _{0}]\\
&  & +[f_{l}(\epsilon )\Gamma _{l}+f_{r}(\epsilon )
\Gamma _{r}]J_{m}(\tilde{u}_{0})J_{n}(\tilde{u}_{0}).
\end{eqnarray*}
To satisfy the condition that the total current flowing through the 
fictitious voltage probe \( \phi  \) vanishes, one gets 
\begin{equation}
\label{FermiF}
f_{\phi }(\epsilon )J_{m}(\tilde{u}_{\phi })J_{n}(\tilde{u}_{\phi })
=\overline{f}(\epsilon )\Gamma _{0}\frac{J_{m}(\tilde{u}_{0})J_{n}
(\tilde{u}_{0})}{i(m-n)\omega _{0}+\Gamma _{0}}\, ,
\end{equation}
where \( \overline{f}(\epsilon )=(f_{l}\Gamma _{l}+f_{r}\Gamma _{r})/
\Gamma _{0} \) is an effective Fermi function of the dot without the
fictitious voltage probe.

Using Eq.(\ref{Ialpha}) and Eq.(\ref{FermiF}), the time-dependent current
flowing through lead \( \alpha  \) can be expressed as

\begin{eqnarray}
I_{\alpha }(t) & = & \frac{e}{\hbar }\Gamma _{\alpha }\int \frac{d\epsilon }
{2\pi }\sum ^{\infty }_{m,n=-\infty }e^{i(m-n)\omega _{0}t}\Phi _{mn}
{\mathcal{F}}^{(\phi )}_{mn}(\epsilon )\nonumber \\
 &  & \times \left\{ i(m-n)\hbar \omega _{0}f_{\alpha }(\epsilon )+
\Gamma _{\bar{\alpha }}\left[ f_{\alpha }(\epsilon )
-f_{\bar{\alpha }}(\epsilon )\right] \right\} ,\label{Ialpha1} 
\end{eqnarray}
where 
\begin{eqnarray*}
{\mathcal{F}}^{(\phi )}_{mn}(\epsilon ) & = & \frac{J_{m}(\tilde{u})J_{n}
(\tilde{u})}{\tilde{\epsilon }^{(\phi )}_{m}
[\tilde{\epsilon }^{(\phi )}_{n}]^{*}}\\
\Phi _{mn} & = & (m-n)\hbar \omega _{0}-\frac{i(\Gamma _{0}
+\Gamma _{\phi })}{(m-n)\hbar \omega _{0}-i\Gamma _{0}},
\end{eqnarray*}
Dephasing effect is represented in \( \Phi _{mn} \)
through \( \Gamma _{\phi} \).

The average current through the quantum dot can be 
obtained by summing the terms with \( m=n \) in Eq.(\ref{Ialpha1}), 
\begin{eqnarray}
\frac{<I>}{I_{0}} & = & \int \frac{d\epsilon }{2\pi }
[f_{l}(\epsilon )-f_{r}(\epsilon )]\nonumber \\
 & \times  & \sum _{m=-\infty }^{\infty }\frac{\Gamma _{0}
+\Gamma _{\phi }}{(\epsilon -\epsilon _{0}-m\hbar \omega _{0} )^{2}
+\left( \frac{\Gamma _{0}+\Gamma _{\phi }}{2}\right) ^{2}},\nonumber 
\end{eqnarray}
where 
\[
I_{0}=\frac{e}{h}\frac{\Gamma _{l}\Gamma _{r}}{\Gamma _{l}+\Gamma _{r}}.\]

\begin{figure}
{\par\centering \resizebox*{0.9\columnwidth}{!}{\includegraphics
{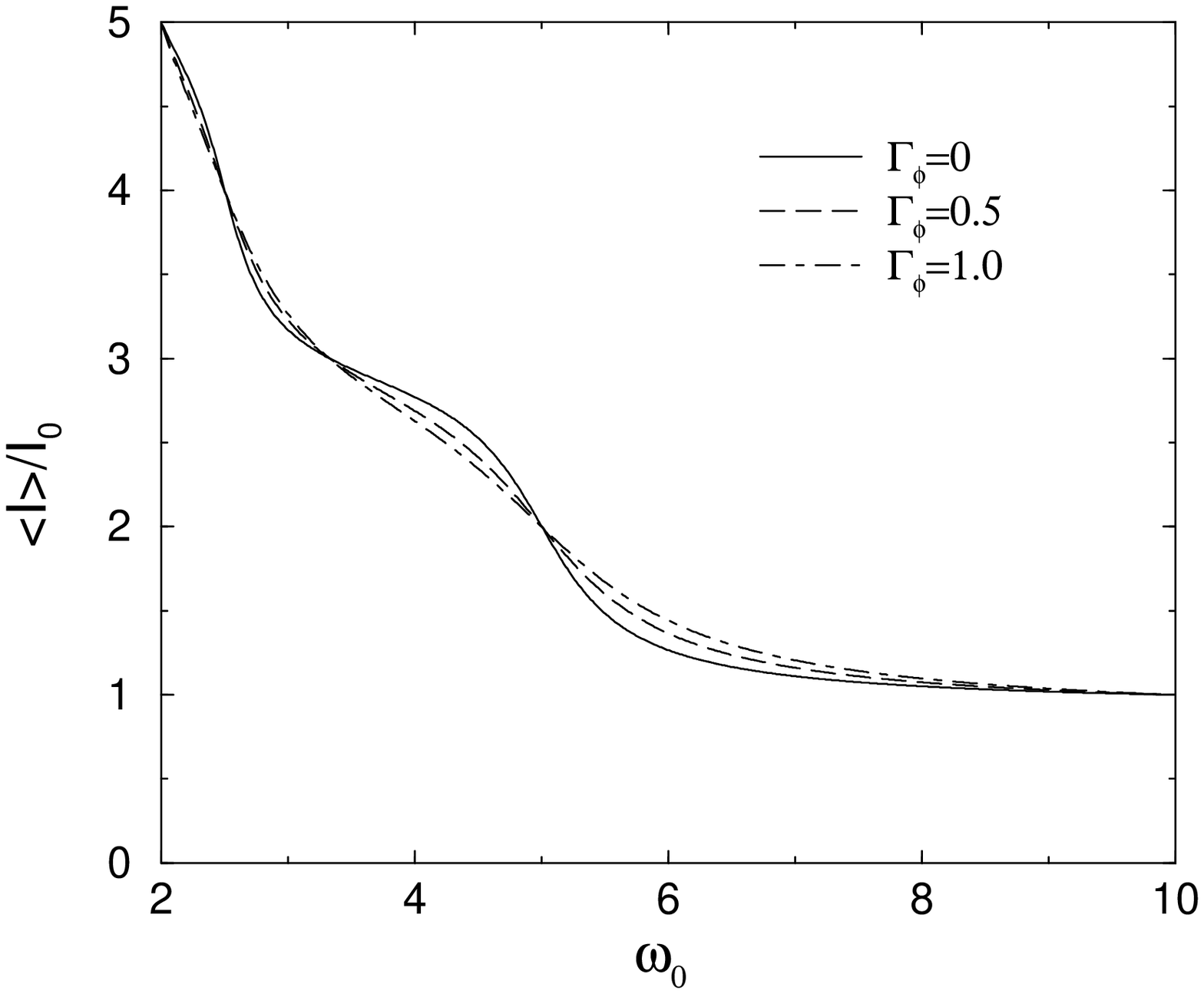}} 
\par}

\caption{The time-averaged current for the single quantum dot
with \protect\( \Gamma _{l}=\Gamma _{r}=0.5\protect \),
\protect\( \epsilon _{0}=0\protect \), and 
\protect\( \mu_{r}=-\mu_{l}=5\protect \).}

\label{fig: 1}
\end{figure}

Figure 1 shows the time-averaged current versus $\omega _{0}$.
It can be clearly seen that the average current density through the quantum
dot consists of a series of the resonant peaks\cite{meir}
at \( \epsilon _{0}+m\hbar \omega  \)
and the dephasing causes the broadening of the resonant peaks.
The dephasing only has a minor effect on the \( I-V \) curve 
since the current amplitude is insensitive to the dephasing. 
In particular, at large $\mu_{R}-\mu_{L}$, $<I>$ approaches $I_{0}$,
independent of $\Gamma_{\phi}$.

\textbf{\underbar{Double Quantum Dots}}. We can generalize the single dot to
a coupled double-dot system.
The Hamiltonian is 
\begin{eqnarray*}
H & = & H_{0}+H_{\phi }\\
H_{0} & = & \sum _{\alpha ,k}\epsilon _{\alpha ,k}(t)
c^{+}_{\alpha ,k}c_{\alpha ,k}+\sum _{\alpha }\epsilon _{\alpha }(t)
d^{+}_{\alpha }d_{\alpha }\\
&  & +\sum _{\alpha ,k}[t_{\alpha k}c^{+}_{\alpha k}d_{\alpha }+h.c.]
+[\Delta d^{+}_{l}d_{r}+h.c.],\\
H_{\phi } & = & \sum _{\alpha ,k}\epsilon ^{(\phi )}_{\alpha ,k}(t)
c^{(\phi )+}_{\alpha ,k}c^{(\phi )}_{\alpha ,k}
+\sum _{\alpha ,k}[t^{(\phi )}_{\alpha k}
c^{(\phi )+}_{\alpha k}d_{\alpha }+h.c.].
\end{eqnarray*}
Comparing with the Hamiltonian for a single dot, two fictitious voltage probes
\( c^{(\phi )}_{\alpha ,k} \) (\( \alpha =l,r \)) 
are connected to the two dots.
The current flowing in and out the reservoirs can 
still be calculated by Eq.(\ref{Ialpha})
provided that \( {\mathbf{G}}^{r} \) and \( {\mathbf{G}}^{<} \) are now
\( 2\times 2 \)
matrices with the matrix elements associated with
the left and right dots. 
The time-dependent current flowing
through the measurement lead $\alpha$ 
(\( \eta =0 \), \( \alpha =l,\, r \)) or the fictitious
voltage lead $\alpha$
(\( \eta =\phi  \), \( \alpha =l,\, r \)) can be expressed as
\begin{eqnarray}
&  & I^{(\eta )}_{\alpha }(t)=-\frac{e}{\hbar }
\Gamma_{\alpha }\int \frac{d\epsilon }{2\pi }
\left[ \sum _{\beta }f^{(\phi )}_{\beta }(\epsilon )
\Gamma^{(\beta )}_{\phi }|
A^{(\phi ,\beta )}_{\alpha \beta }(\epsilon ,t)|^{2}\right. \nonumber \\
&  & \left. +\sum _{\beta }f_{\beta }(\epsilon )\Gamma _{\beta }
|A^{(\beta )}_{\alpha \beta }(\epsilon ,t)|^{2}+2f_{\alpha }(\epsilon )
{\mathrm{Im}}A^{(\eta ,\alpha )}_{\alpha \alpha }(\epsilon ,t)\right],
\label{Ialpha2} 
\end{eqnarray}
where \( A^{(\eta ,\beta )}_{\alpha \beta }(t,\epsilon ) \) 
(\( \eta =0,\phi  \))
are the matrix elements of the spectral function. The diagonal elements are 
\[A^{(\eta ,\alpha )}_{\alpha \alpha }(\epsilon ,t)=
e^{-i\tilde{u}^{(\eta )}_{\alpha }\sin \omega _{0}t}
\sum^{\infty }_{m=-\infty }\frac{J_{m}({\tilde{u}}^{(\eta )}_{\alpha })
e^{im\omega _{0}t}}{{\mathcal{E}}^{(\phi )}_{\alpha }(\epsilon ,m)},\]
and the non-diagonal elements (\( \alpha \neq \beta  \)) are 
\begin{eqnarray}
A^{(\eta ,\beta )}_{\alpha \beta }(\epsilon ,t) & = & 
\Delta e^{-i(\tilde{u}^{(\eta )}_{\beta }
+\tilde{u}_{0})\sin \omega _{0}t}\nonumber \\
& \times  & \sum ^{\infty }_{m=-\infty }\frac{J_{m}
(\tilde{u}^{(\eta )}_{\beta })
e^{im\omega _{0}t}}{{\mathcal{E}}^{(\phi )}_{\beta }(\epsilon ,m)}
\sum^{\infty }_{k=-\infty }\frac{J_{k}(\tilde{u}_{0})e^{ik\omega _{0}t}}
{\tilde{\epsilon }^{(\phi )}_{\alpha }(m+k)},\nonumber 
\end{eqnarray}
with 
\begin{eqnarray*}
{\mathcal{E}}^{(\alpha )}_{m}(\epsilon ,\phi ) & = 
& \tilde{\epsilon }^{(\phi )}_{\alpha }(m)-|\Delta |^{2}
\sum _{k}\frac{J^{2}_{k}(\tilde{u}_{0})}
{\tilde{\epsilon }^{(\phi )}_{\bar{\alpha }}(m+k)},\\
\tilde{\epsilon }^{(\phi )}_{\alpha }(m) & = 
& \epsilon -\epsilon _{\alpha }-m\hbar \omega _{0}
+i\frac{\Gamma _{\alpha }+\Gamma ^{(\alpha )}_{\phi }}{2},\\
\tilde{u}_{\alpha } & = & \frac{u_{\alpha }-v_{\alpha }}{\hbar \omega _{0}},\\
\tilde{u}^{(\phi )}_{\alpha } & = 
& \frac{u_{\alpha }-v^{(\phi )}_{\alpha }}{\hbar \omega _{0}},\\
\tilde{u}_{0} & = & \frac{u_{1}-u_{2}}{\hbar \omega _{0}}\, .
\end{eqnarray*}
We further require the net current through each fictitious probe to 
vanish, \textit{i.e.}, \( I^{(\phi )}_{\alpha }(t)=0 \). 
From this constraint, we can obtain
relations among \( f^{(\phi )}_{\beta }(\epsilon ) \), 
\( f_{\beta }(\epsilon ) \),
\( A^{(\phi ,\beta )}_{\alpha ,\beta }(\epsilon ,t) \) and
\( A^{(\beta )}_{\alpha ,\beta }(\epsilon ,t) \). Using these relations to
replace \( \sum _{\beta }f^{(\phi )}_{\beta }(\epsilon )
\Gamma ^{(\beta )}_{\phi }|A^{(\phi ,\beta )}_{\alpha \beta }
(\epsilon ,t)|^{2} \)
in Eq.(\ref{Ialpha2}), we obtain the current through lead \( \alpha  \),
\begin{eqnarray}
I_{\alpha }(t) & = & \frac{e}{\hbar }\int \frac{d\epsilon }{2\pi }
\sum^{\infty }_{m,n=-\infty}e^{i(m-n)\omega _{0}t}
\Phi _{mn}(\epsilon)\Gamma _{\alpha }\nonumber \\
&  & [i(m-n)\hbar \omega _{0}f_{\alpha }(\epsilon )
{\mathcal{Q}}^{(\alpha )}_{mn}(\epsilon ,\phi )+\Gamma _{\bar{\alpha }}
|\Delta |^{2}{\mathcal{I}}_{\alpha }(\epsilon )].
\end{eqnarray}
In the above equation,
\begin{eqnarray*}
{\mathcal{I}}^{(j)}_{\alpha }(\epsilon ) & = 
& f_{\alpha }(\epsilon ){\mathcal{F}}^{(\alpha )}_{mn}
(\epsilon ,\phi )-f_{\bar{\alpha }}(\epsilon )
{\mathcal{F}}^{(\bar{\alpha })}_{mn}(\epsilon ,\phi ),\\
{\mathcal{F}}^{(\alpha )}_{mn}(\epsilon ,\phi ) & = 
& \frac{J_{m}(\tilde{u}_{\alpha })J_{n}(\tilde{u}_{\alpha })}
{{\mathcal{E}}^{(\alpha )}_{m}(\epsilon ,\phi )
{\mathcal{E}}^{(\alpha )*}_{n}(\epsilon ,\phi )}
\Lambda ^{(\overline{\alpha })}_{mn}(\epsilon ,\phi ),\\
{\mathcal{Q}}^{(\alpha ,\phi )}_{mn}(\epsilon ) & = 
& \frac{J_{m}(\tilde{u}_{\alpha })J_{n}(\tilde{u}_{\alpha })}
{{\mathcal{E}}^{(\alpha )}_{m}(\epsilon ,\phi )
{\mathcal{E}}^{(\alpha )*}_{n}(\epsilon ,\phi )}
\left[ 1+|\Delta |^{2}\Lambda ^{(\bar{\alpha })}_{mn}
(\epsilon ,\phi )\right. \\
&  & \left. +\Gamma ^{(\bar{\alpha })}_{\phi }\frac{1-|\Delta |^{4}
\Lambda ^{(\alpha )}_{mn}(\epsilon ,\phi )\Lambda ^{(\bar{\alpha })}_{mn}
(\epsilon ,\phi )}{\Omega ^{(\bar{\alpha })}_{2,mn}(\epsilon )}\right] \\
\Lambda ^{(\alpha )}_{mn}(\epsilon ,\phi ) & = & \sum ^{\infty }_{k
=-\infty }\frac{J^{2}_{k}(\tilde{u}_{0})}
{\tilde{\epsilon }^{(\phi )}_{\alpha }(m+k)
[\tilde{\epsilon }^{(\phi )}_{\alpha }(n+k)]^{*}}\\
\Phi _{mn}(\epsilon ) & = & \frac{\Omega ^{(l)}_{1,mn}(\epsilon )
\Omega ^{(r)}_{1,mn}(\epsilon )/\Omega ^{(l)}_{0,mn}(\epsilon )
\Omega ^{(r)}_{0,mn}
(\epsilon )}{1-|\Delta |^{4}\Gamma ^{(l)}_{\phi }\Gamma ^{(r)}_{\phi }
\frac{\Lambda ^{(l)}_{mn}(\epsilon ,\phi )\Lambda ^{(r)}_{mn}(\epsilon ,
\phi )}{\Omega ^{(l)}_{0,mn}(\epsilon )\Omega ^{(r)}_{0,mn}(\epsilon )}}\\
\Omega ^{(\alpha )}_{j,mn}(\epsilon )& = & i(m-n)\omega _{0}-
\left( \Gamma _{\alpha }+j\Gamma ^{(\alpha )}_{\phi }\right) +\\
 &  & |\Delta |^{2}\left[ i(m-n)\omega _{0}-\left( \Gamma _{\bar{\alpha }}
+\Gamma ^{({\bar{\alpha }})}_{\phi }\right) \right] 
\Lambda^{(\alpha )}_{mn}(\epsilon ,\phi )
\end{eqnarray*}

The time average of the current 
\( \overline{I}=\left\langle I_{l}(t)\right\rangle _{t}=
-\left\langle I_{r}(t)\right\rangle _{t} \)
can be obtained 
\begin{eqnarray}
\overline{I} & = & \frac{e}{\hbar }\Gamma _{l}\Gamma _{r}|\Delta |^{2}
\int \frac{d\epsilon }{2\pi }\sum ^{\infty }_{m=-\infty }
\Phi _{mm}(\epsilon )\nonumber \\
&  & [f_{l}(\epsilon ){\mathcal{F}}^{(l)}_{mm}
(\epsilon ,\phi )-f_{r}(\epsilon ){\mathcal{F}}^{(r)}_{mm}
(\epsilon ,\phi )].\label{Ibar2} 
\end{eqnarray}
From the above expression, one finds a non-zero tunneling 
current even without
{\it dc} bias between the right and left leads. It is consistent with the 
experimental observations\cite{oosterkamp,schedelbeck,tsui}. 

\begin{figure}
{\par\centering \resizebox*{0.9\columnwidth}{!}{\includegraphics
{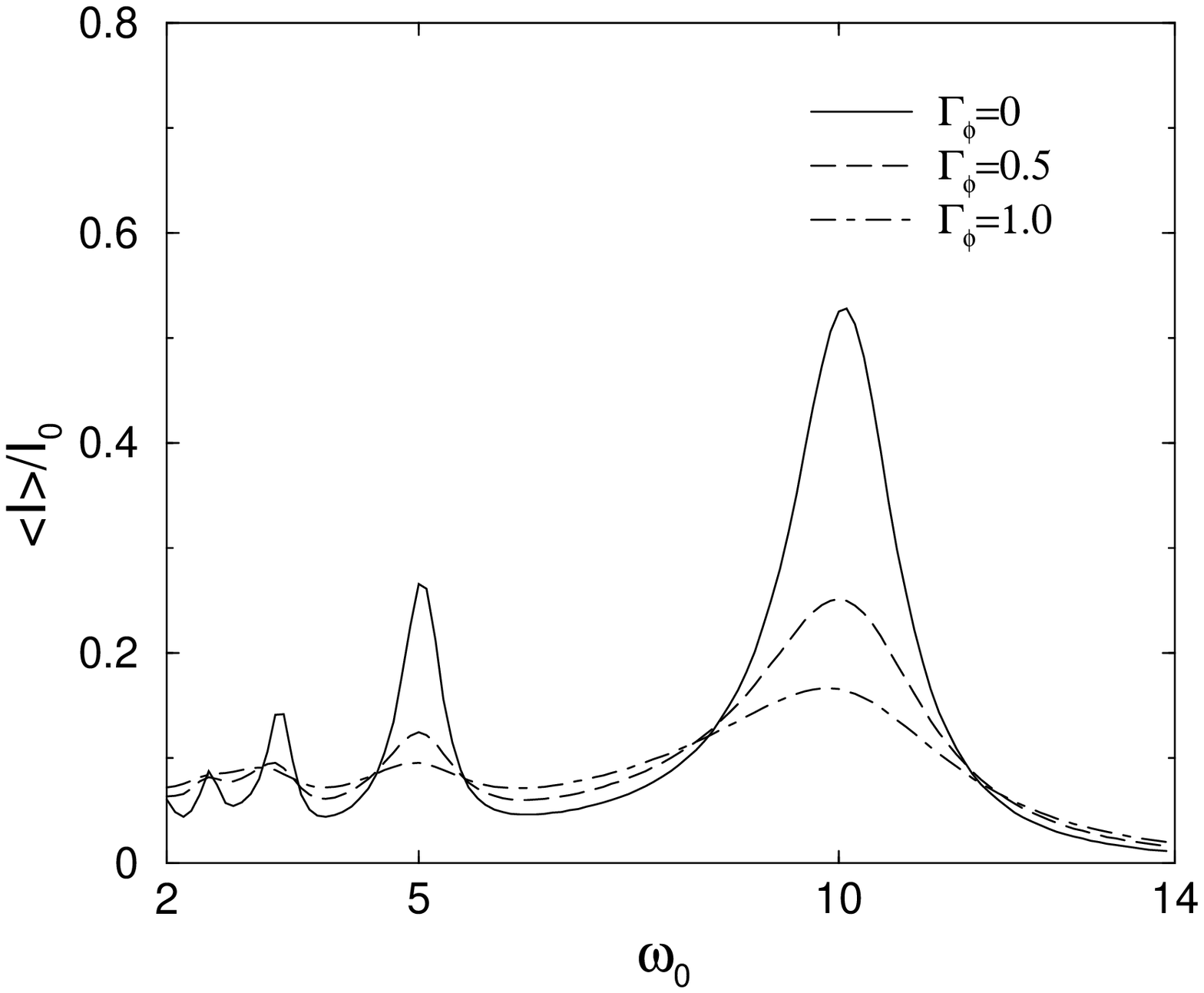}}
\par}
 
\caption{The time-averaged current for the double dots with
the parameters: \protect\( \Gamma _{l}=\Gamma _{r}=0.5\protect \),
\protect\( \epsilon _{r}-\epsilon _{l}=10\protect \),
\protect\( \mu _{l}=\mu _{r}=0\protect \), \protect\( v_{ac}=6\protect \)
and \protect\( T=0\protect \).
\label{fig: 2}}
\end{figure}

Figure \ref{fig: 2} shows the time-averaged tunneling current $<I>$ for 
different dephasing strength
\( \Gamma _{\phi } \). $<I>$ strongly depends on \( \Gamma _{\phi } \).
In real systems, \( \Gamma _{\phi } \) is a function of
temperature. Temperature has two effects on the tunneling
current: the direct contribution from Fermi distribution
function and the indirect contribution through
\( \Gamma _{\phi } \). 
We find that at low temperatures the temperature dependence
of the current is essentially determined by the temperature dependence
of \( \Gamma _{\phi} \). The current is almost independent of temperature
within \( kT\ll |\epsilon _{l}-\epsilon _{r}| \) while the dephasing strength
\( \Gamma _{\phi } \) is set to a constant. 
Thus, by measuring the temperature dependence
of the current, it is possible to determine the temperature dependence of the
dephasing rate in double-dot systems.

\begin{figure}
{\par\centering \resizebox*{0.9\columnwidth}{!}{\includegraphics
{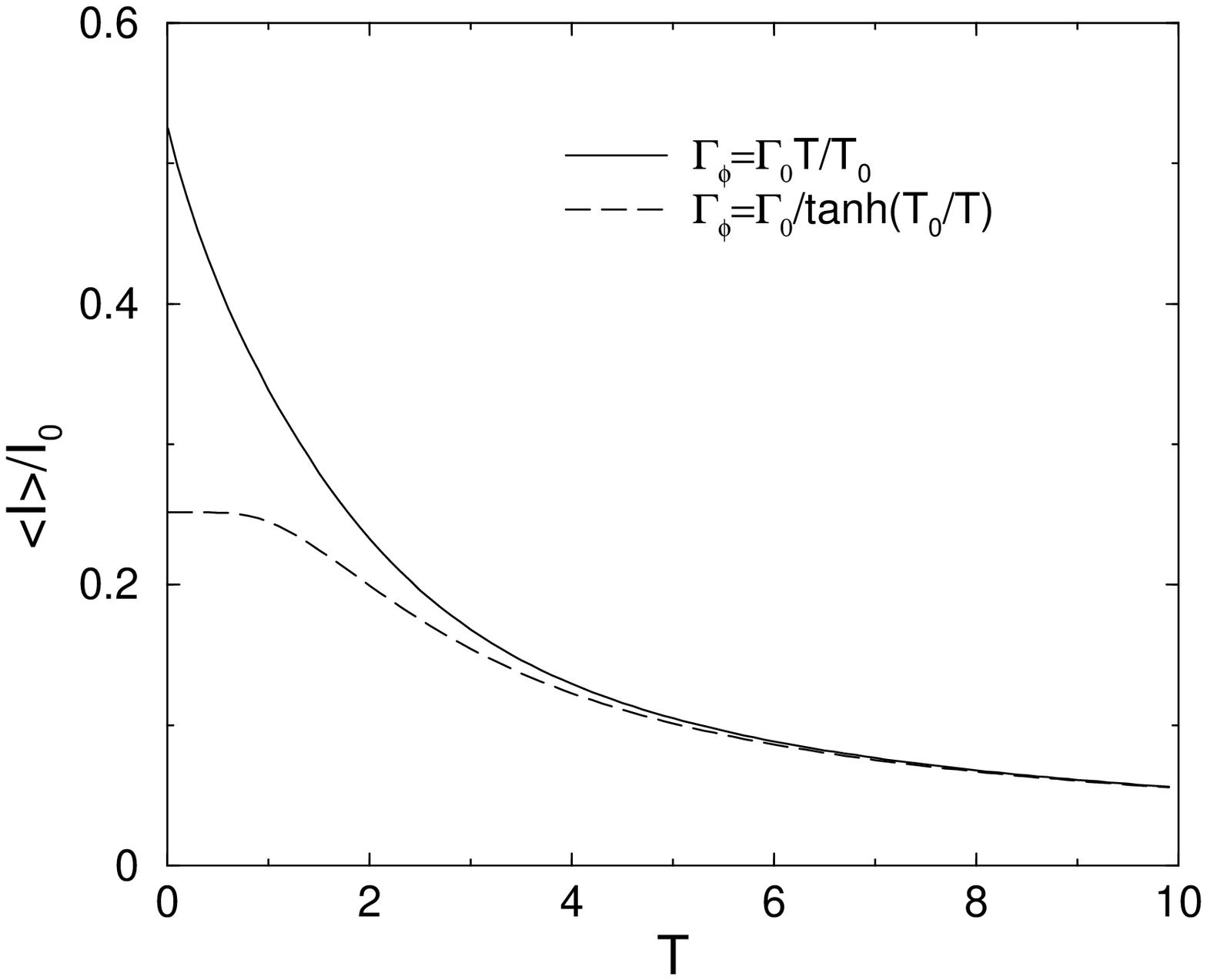}}
\par}
 
\caption{The temperature dependence of the average current.
\protect\( \omega _{0}=10\protect \) and
the parameters are the same
as in Fig.\ref{fig: 2}.
The dephasing strength \protect\( \Gamma_{\phi }\protect \)
has two different kinds of temperature dependence as discussed in the text.
\label{fig: 3}}
\end{figure}

It is commonly believed that the dephasing strength \( \Gamma _{\phi } \)
is a power law function of temperature $T$. Hence,
\( \Gamma _{\phi } \)  approaches zero as $T$ goes to zero.
However, this consensus has been challenged in a recent 
experiment\cite{Mohanty} in which \( \Gamma _{\phi } \) was
found to saturate to finite values in many one or two dimensional
systems. The question remains whether such a saturation 
exists in the zero-dimensional
quantum dots. We suggest that this issue can be resolved in
double-dot systems by studying the temperature dependence of the
tunneling current. 
Figure \ref{fig: 3} shows the temperature dependence of the average current. The
current is calculated by assuming two different kinds of
temperature dependence of
the dephasing strength \( \Gamma _{\phi } \) \cite{Mohanty}. The first
kind is the normal linear dependence
(Other power-law dependence
gives rise to qualitatively similar results.)
with \( \Gamma_{\phi }=\Gamma _{0}\frac{T}{T_{0}} \)
($\Gamma _{0}=0.5$ and $T_{0}=2$ in the plot.).
The second kind is the
abnormal temperature dependence with 
\( \Gamma _{\phi }=\Gamma _{0}/\tanh (\frac{T_{0}}{T}) \), in which the
dephasing rate saturates at low temperatures. 
Figure 3 clearly shows that the average current exhibits distinct
behaviors at low $T$ for these 
two temperature dependence of \( \Gamma _{\phi } \).
Thus, one can detect the possible dephasing rate saturation 
in a double-dot system by measuring the tunneling current.

Before summary, we would like to comment on the difference
between single-dot and double-dot systems. Although the
dephasing effect shows up in both single-dot and double-dot
systems, the most pronounced effect is under the pumping
situation (Fig.2) in double dots (There is no pumping effect in single-dot
systems.). In a non-pumping situation in double
dots, i.e., the chemical potentials
are not equal on the left and the right leads, the dephasing effect
is reduced. To see this, one can rewrite the current (Eq.(8))
into two terms, one contains $(f_{l}(\epsilon)-f_{r}(\epsilon))
({\mathcal{F}}^{(l)}_{mm}
(\epsilon ,\phi )+{\mathcal{F}}^{(r)}_{mm}(\epsilon ,\phi ))$
and the other contains $(f_{l}(\epsilon)+f_{r}(\epsilon))
({\mathcal{F}}^{(l)}_{mm}
(\epsilon ,\phi )-{\mathcal{F}}^{(r)}_{mm}(\epsilon ,\phi ))$. 
The first term shows a similar behavior for the current
in a single-dot, whereas the second term gives a similar behavior 
for the pumping current. Thus, in a non-pumping set-up,
the tunneling current is like the
pumping case (Fig.2) superimposed by a curve similar as shown
in Fig.1 which makes the dephasing effect less pronounced.

In summary, we analyze the dephasing effects in photon-assisted
tunneling in quantum dots. The dephasing effect is introduced by
using fictitious voltage probes. 
We find that the time-averaged current is insensitive to
dephasing in single-dot systems. However, in the pumping
set-up of double-dot systems,
dephasing has profound effect in the tunneling current which can be
measured to determine the dephasing rate.

\noindent \textbf{Acknowledgments:} 
This work was supported by US-DOE and NSF-China.

\end{document}